# Design and Simulation of a Hybrid Controller for a Multi-Input Multi-Output Magnetic Suspension System

Sherif M. Abuelenin, *Member, IEEE*

*Abstract*—In this paper we present a Fuzzy Logic control approach designed to stabilize a multi-input multi-output magnetic suspension system. The system has four cubic floaters and four actuators that apply magnetic forces on the floaters, the suspension is performed by changing the voltages applied on the actuators, hence changing their currents, producing vertical magnetic forces that balance with the gravitational force. A fuzzy logic controller is used to control each actuator; the system is nonlinear and sensitive to initial conditions. Another fuzzy logic controller is used as a supervisory controller in order to increase the dynamic range of the system, enabling it to stabilize the floaters when the initial displacements are relatively big. Another design consideration was to keep the four floaters in the same plane as much as possible, to perform that task, a PD controller was set to modulate the currents of the four actuators in order to minimize an error signal measuring the relative vertical displacement of all the four floaters. Simulation results show that the designed control scheme stabilized the system for the design constrains.

## I. INTRODUCTION

Magnetic suspension systems are systems in which a rotor or a floater is suspended in magnetic field without contact with the surroundings. The open-loop system model examined in this paper was introduced in [1]. Magnetic Suspension is accomplished in these systems through automatic control of actuators currents by changing their currents. The magnetic force created by the actuators acts upon the floater in the opposite direction of gravity to keep it suspended. Feedback of position information of the floater is required to create a closed-loop system and is provided either through sensors or by using self-sensing methods [2]-[5].

Magnetic suspension systems are being increasingly used in many applications, including industry, since they are wear-free [1]-[5]. These systems are highly nonlinear and highly dependant on initial condition, using linear controllers can produce the desired dynamic response only for the region in which a linear model of the system was created. Nonlinear control provides the ability to create desired dynamic responses, many nonlinear control algorithms were introduced in research [2]-[7].

In this paper, a fuzzy logic controller is introduced to control a multi-input multi-output magnetic suspension system that was formulated in [1]. The system has four cubic shaped floaters with equal masses and sizes, but the four actuators have different electromagnetic properties.

Fig. 1 shows a cross-section of the magnetic suspension system model to be controlled; there are four equal–size cubic electromagnetic actuators that are securely attached to the four corners of a rectangle plate stator mounted on a stationary metal plate. The DC current in the coil of the $k$-th actuator $i_k(t)$, (where $k = 1,2,3,4$) can be adjusted by changing the DC voltage $u_k(t)$ applied to the coil wrapped around an iron core. This will result in varying magnetic force $f_k(t)$ acting vertically on iron floater $k$. The top view schematic illustrated in Fig. 2 shows the four cubic floaters, which are equal in size and mass.

The mass density of the floaters is assumed to be uniform. There are only two forces acting upon each of the floaters – the magnetic force and the gravity. Through the balancing of these two vertical forces, the floaters remain suspended in the air. The distance between the $k$-th floater and the $k$-th actuator, $z_k(t)$, is measured in real-time by distance sensor $k$. The origin of the z axis is marked in Fig. 1, so is the target horizontal level of the floaters.

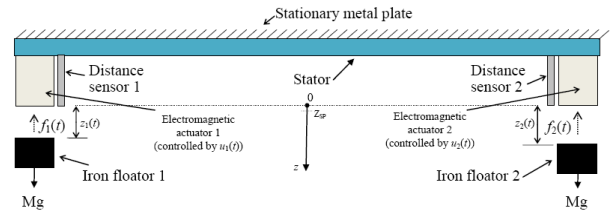

Fig. 1. Cross-section view of the simplified magnetic suspension system

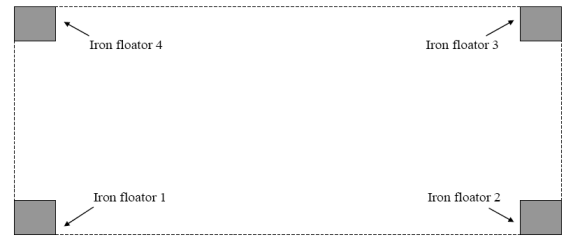

Fig. 2. Aerial view of the four equal-size iron floaters that are aligned horizontally and vertically. An electromagnetic actuator is placed above each floater (not shown here; see Fig. 1 for actuators 1 and 2)

S. M. Abuelenin is with the Department of Electrical Engineering, Sinai University, Arish, Egypt. Phone: +20-10-536-5313; (e-mail: sherif217@ieee.org).

The control approach introduced is based on incorporating two fuzzy controllers for each actuator (electromagnet); a main one to control the actuator current, and another (supervisory controller) to tune the output gain of the first controller. A total of 8 Fuzzy controllers are incorporated in the design. A Proportional-Derivative (PD) controller is inserted to monitor the relative planar displacement of the four floaters and minimize this displacement as much as possible.

The whole system is simulated in SIMULINK with the Fuzzy controllers implemented using MATLAB Fuzzy Logic toolbox.

## II. SYSTEM MODEL

### A. Mathematical Model

The behavior of each of the four actuator-floater pairs is governed by the following equations [1]:

$$R_k \cdot i_k(t) + \frac{A_k N_k^2 \mu_0}{2 z_k(t)} \cdot \frac{d i_k(t)}{dt} - \frac{A_k N_k^2 \mu_0 i_k(t)}{2 z_k^2(t)} \cdot \frac{d z_k(t)}{dt} = u_k(t) \quad (1)$$

$$f_k(t) = \frac{A_k N_k^2 \mu_0}{4} \left( \frac{i_k(t)}{z_k(t)} \right)^2 \quad (2)$$

$$M \frac{d^2 z_k(t)}{dt^2} = f_k(t) - Mg \quad (3)$$

The meanings, values (and their limits) and units of the parameters of equations (1), (2), and (3) are given in Table I.

There are four pairs of the equations characterizing the four electromagnetic actuators. The four actuators are similar in size and mass but have different electromagnetic properties.

## III. CONTROL DESIGN

Developing the control approach discussed here was performed in three steps, resulting in a controller that is a combination of fuzzy controllers, supervised by another set of fuzzy controllers, and a Proportional-Derivative controller.

Fuzzy control is a convenient alternative method of nonlinear control used for a variety applications since it provides a method for constructing control algorithms via the use of heuristic information that represent the "rules" according to which we would like the controlled process to perform [8].

In our approach, first, a main fuzzy logic controller (FLC$_k$) is used for each of the four actuators to control the voltage applied to it $u_k(t)$, one of the inputs to this FLC is the error signal (the difference between the reference input set-point and the position of the floater), that should equal zero in steady-state, the other input is the derivative of the error signal.

TABLE I
SYMBOLS AND THEIR MEANINGS AND VALUES (K = 1,2,3,4).

| Symbol | Meaning | Value, Limit, and Unit |
|---|---|---|
| $f_k(t)$ | Magnetic force acting up actuator k | >=0, N |
| $z_k(t)$ | Distance between actuator k and floater k | >0, m |
| $\mu_0$ | Magnetic conductivity in the air | $4\pi*1e-7$ H/m |
| M | Mass of each floater | 3kg |
| g | Acceleration due to gravity | 9.8m/s$^2$ |
| $u_k(t)$ | DC control voltage of actuator k | 0-50V, k=1,2<br>0-70V, k=3<br>0-60V, k=4 |
| $R_k$ | Coil resistance of coil k | 5Ω, k=1,2<br>10Ω, k=3<br>8.5Ω, k=4 |
| $i_k(t)$ | DC control current of actuator k | 0-10A, k=1,2<br>0-7A, k=3,4 |
| $A_k(t)$ | Sectional area of actuator k | 0.0002m$^2$, k=1<br>0.000237m$^2$, k=2<br>0.0005m$^2$, k=3<br>0.0004m$^2$, k=4 |
| $N_k$ | Coil loop number of actuator k | 300, k=1,2<br>600, k=3<br>500, k=4 |

Gains are introduced on both inputs and on the output of the controller to allow tuning of the controller, tuning of both input gains were done manually for each actuator. The closed-loop system is nonlinear and sensitive to initial condition; hence, the second step was to introduce another fuzzy controller that is used as a supervisor to tune the output gain of the first FLC in order to achieve stability for a wider range of initial conditions. Each supervisory FLC (SFLC) has one input, which is the error signal, and its output is the value of the output gain for the corresponding main FLC.

The third step was the introduction of a control method to level the four floaters, as one of the design criteria was to keep the four floaters in the same plane all the times, which means they should satisfy the constraint $z_1(t) + z_3(t) = z_2(t) + z_4(t)$ as much as possible. In order to satisfy this, a Proportional-Derivative (PD) controller is used to modulate the voltages applied to the four actuators, by adding to (or subtracting from) a value determined by the controller. The input signal to the PD controller is the level error signal; $l(t) = z_1(t) + z_3(t) - z_2(t) - z_4(t)$.

### A. The Main Fuzzy Controller

In each of the main FLCs, Five linguistic variables are used to describe each Fuzzy input and output variable. Fig. 3

shows the linguistic variables and their associated membership functions used in FLC1 (used to control $u_1(t)$).

The "rule-base" used in this controller is summarized in Table II, and the resulting control surface is shown in Fig. 4

TABLE II
SUMMARY OF FUZZY RULES OF THE MAIN FLC

| U(t) | Error membership function (mf) | | | | |
|---|---|---|---|---|---|
| Change of error membership function (mf) | mf1 | mf2 | mf3 | mf4 | mf5 |
| mf1 | mf1 | mf1 | mf1 | mf2 | mf3 |
| mf2 | mf1 | mf1 | mf2 | mf3 | mf4 |
| mf3 | mf1 | mf2 | mf3 | mf4 | mf5 |
| mf4 | mf2 | mf3 | mf4 | mf5 | mf5 |
| mf5 | mf3 | mf4 | mf5 | mf5 | mf5 |

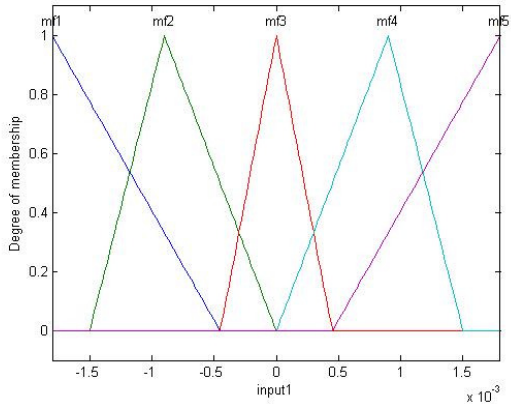

Fig. 3.a. Membership functions for "input 1; the error" variable of the main FLC of actuator 1

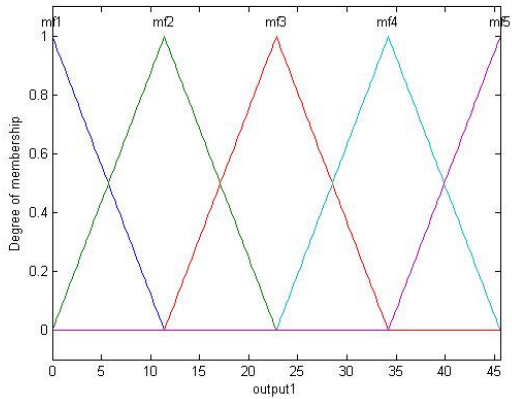

Fig. 3.b. Membership functions for "output" variable of the main FLC of actuator 1

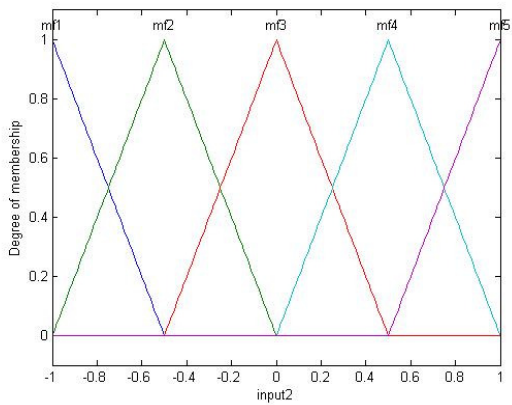

Fig. 3.c. Membership functions for "input 2; the error rate" variable of the main FLC of actuator 1

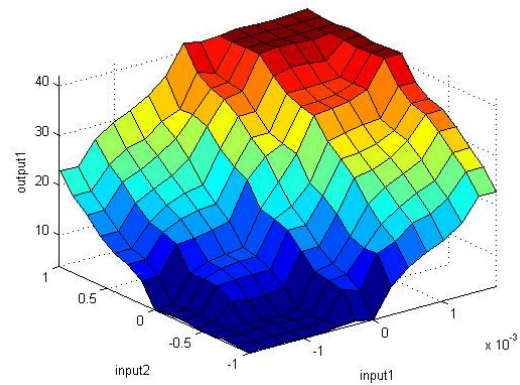

Fig. 4. Fuzzy Control Surface of the main FLC for electromagnet 1

The other three controllers have the similar variables and membership functions, except that each is designed to provide the steady state value of $u_k(t)$ required for zero-error steady state.

### B. The Supervisory Fuzzy Controller

The function of each SFLC is to tune the output gain of the corresponding main FLC in order to enable system stability for a wider range of initial condition. The SFLC is a simple 1-input 1-output FLC which outputs a tuning gain varying from value of 1 (indicating no tuning) to a much higher value determined for each controller. This tuning gain is then multiplied by the output signal of the tuned FLC to produce a modulated control signal. In each SFLC, three linguistic variables are used to describe the input and two to describe the output. Fig. 5 shows the resulting control surface for SFLC1.

### C. The Proportional-Derivative Controller

Simulating the system with the designed Fuzzy Controllers showed that the system stabilized, but the level error signal $l(t)$ was in the order of millimeters. In order to minimize this signal, a PD controller was used to monitor $l(t)$; its output is added (or subtracted) from the control voltage applied to the four actuators simultaneously.

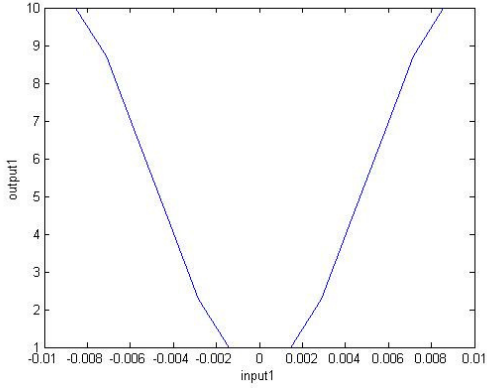

Fig. 5. Fuzzy Control Surface of the Supervisory FLC for electromagnet 1

The output of the PD controller is determined by the following expression: $w(t) = K_p \cdot l(t) + K_d \dfrac{dl(t)}{dt}$, where gains are tuned manually.

## IV. SIMULATION

Fig. 6 shows the SIMULINK model of each open loop actuator system, Fig. 7 shows the same closed loop model with the both the main and the SFLC added. And Fig. 8 shows the model of the complete system with the PD controller. Table III shows the three initial position settings used in the simulation.

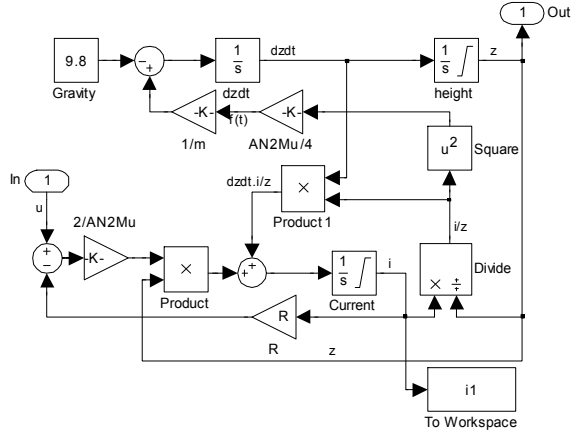

Fig. 6. SIMULINK model of the open loop electromagnet system

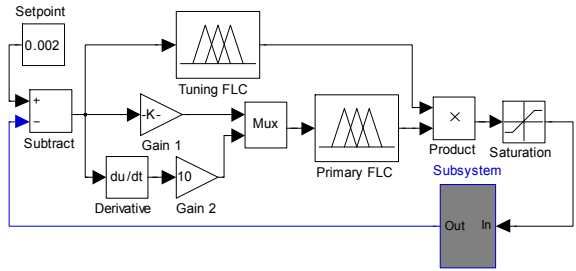

Fig. 7. SIMULINK model of an actuator with both FLCs

TABLE III
INITIAL POSITION SETTINGS

|  | Setting 1 | Setting 2 | Setting 3 |
|---|---|---|---|
| $z_{10}$ | 1mm | 5mm | 6mm |
| $z_{20}$ | 3mm | 3mm | 8mm |
| $z_{30}$ | 9mm | 11mm | 14mm |
| $z_{40}$ | 7mm | 13mm | 12mm |

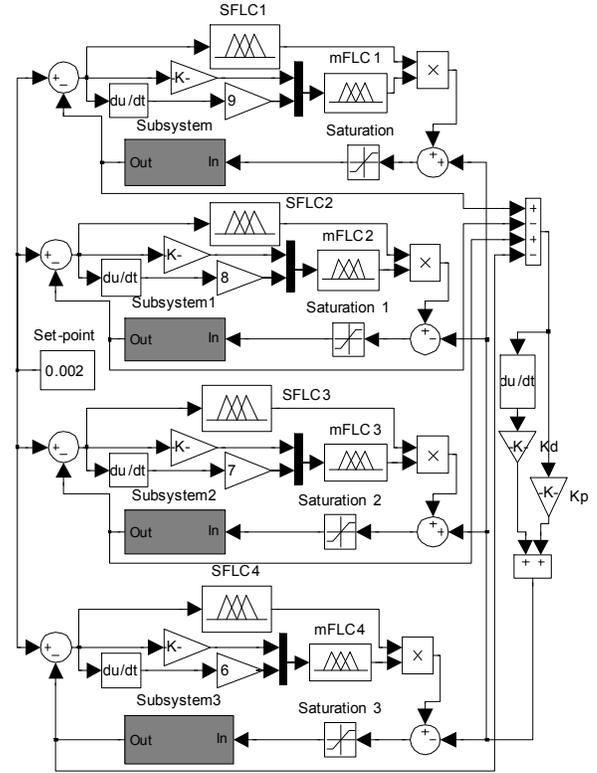

Fig. 8. SIMULINK model of the complete system

Fig. 9.a, b show the position of the four floaters $z_k(t)$ as a function of time as the result of simulating the system with initial position settings 1, before adding the PD controller, and the level error signal $l(t)$. And Fig. 10.a, b show the same plots after the implementation of the PD controller, we see that the PD controller greatly reduced the level error signal $l(t)$ on the expense of increasing the rise time of $z_k(t)$.

Summery of the results for all initial position settings is shown in tables IV-VI.

## V. CONCLUSION

A hybrid nonlinear controller of a multi-input multi-output magnetic suspension system is established by using a combination of four sets of fuzzy supervised fuzzy logic controllers, and a PD controller that is used to tune the outputs of the fuzzy controllers. Simulation of the system showed that the designed control scheme was successful in controlling the vertical position of the four floaters to a pre-specified set-point from different sets of initial positions. Simulation also showed that the addition of the PD controller resulted in minimizing the planar error between the four floaters, keeping them much closer to being in the same plane without much degradation on the performance of each individual floater.

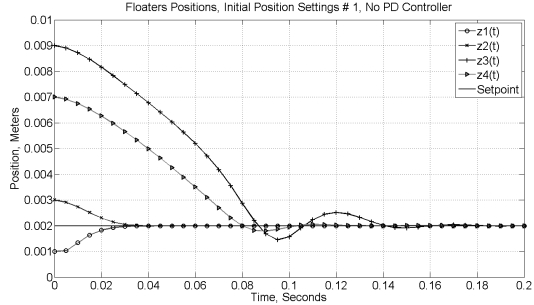

Fig. 9.a. $z_k(t)$ for Initial position settings 1, no PD controller

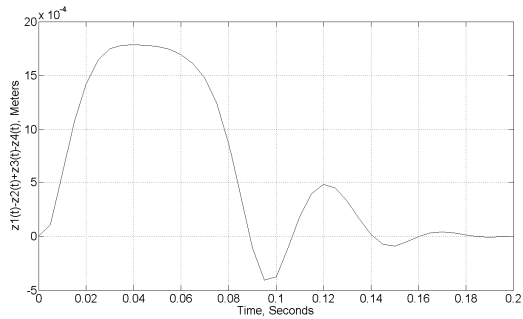

Fig. 9.b. Level difference, Initial position settings 1, no PD controller

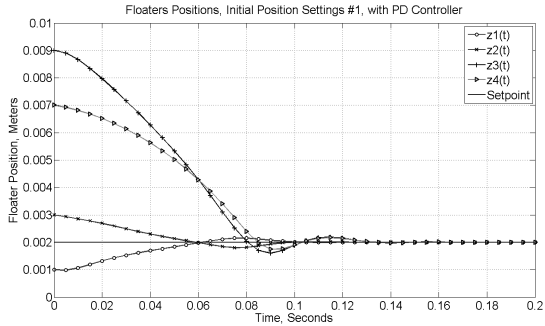

Fig. 10.a. $z_k(t)$ for Initial position settings 1

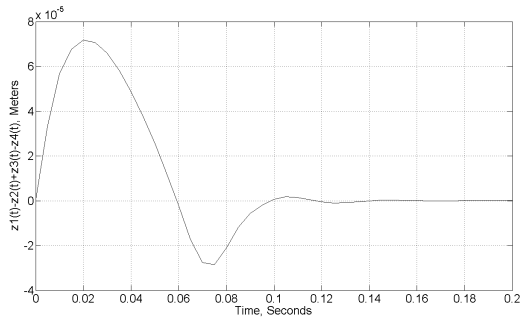

Fig. 10.b. Level difference, Initial position settings 1, (note the scale difference compared to Fig. 9.b.)

TABLE IV
RESULTS FOR INITIAL POSITION SETTING 1

| Initial Position settings 1 | $z_1(t)$ | $z_2(t)$ | $z_3(t)$ | $z_4(t)$ |
|---|---|---|---|---|
| Rise Time, s | 6.200E-02 | 5.950E-02 | 8.050E-02 | 8.450E-02 |
| Max. Abs. Overshoot, m | 2.161E-03 | 1.799E-03 | 1.599E-03 | 1.728E-03 |
| Settling Time, s | 9.900E-02 | 1.130E-01 | 1.210E-01 | 1.240E-01 |
| | Largest Abs. Value | Mean | Std. Deviation | |
| $z_1(t) + z_3(t) - z_2(t) - z_4(t)$ | 7.20E-05 | 1.08E-05 | 2.69E-05 | |

TABLE V
RESULTS FOR INITIAL POSITION SETTING 2

| Initial Position settings 1 | $z_1(t)$ | $z_2(t)$ | $z_3(t)$ | $z_4(t)$ |
|---|---|---|---|---|
| Rise Time, s | 4.650E-02 | 9.050E-02 | 8.550E-02 | 8.050E-02 |
| Max. Abs. Overshoot, m | 1.321E-03 | 2.779E-03 | 1.599E-03 | 1.604E-03 |
| Settling Time, s | 8.650E-02 | 1.025E-01 | 1.245E-01 | 1.205E-01 |
| | Largest Abs. Value | Mean | Std. Deviation | |
| $z_1(t) + z_3(t) - z_2(t) - z_4(t)$ | 1.76E-04 | 5.09E-05 | 5.97E-05 | |

TABLE VI
RESULTS FOR INITIAL POSITION SETTING 3

| Initial Position settings 1 | $z_1(t)$ | $z_2(t)$ | $z_3(t)$ | $z_4(t)$ |
|---|---|---|---|---|
| Rise Time, s | 4.350E-02 | 7.000E-02 | 7.650E-02 | 7.700E-02 |
| Max. Abs. Overshoot, m | 1.191E-03 | 1.879E-03 | 1.274E-03 | 1.430E-03 |
| Settling Time, s | 8.850E-02 | 8.100E-02 | 1.255E-01 | 1.260E-01 |
| | Largest Abs. Value | Mean | Std. Deviation | |
| $z_1(t) + z_3(t) - z_2(t) - z_4(t)$ | 3.69E-04 | 7.25E-05 | 1.06E-04 | |